\def\year{2020}\relax
\documentclass[letterpaper]{article} 
\usepackage{aaai20}  
\usepackage{times}  
\usepackage{helvet} 
\usepackage{courier}  
\usepackage[hyphens]{url}  
\usepackage{graphicx} 
\usepackage{graphicx}  
\usepackage{natbib}
\usepackage{hyperref}
\usepackage{nameref}
\usepackage{titleref}

\usepackage{titlesec}

\usepackage[utf8]{inputenc} 
\usepackage[T1]{fontenc}    
\usepackage{hyperref}       
\usepackage{url}            
\usepackage{booktabs}       
\usepackage{amsfonts}       
\usepackage{nicefrac}       
\usepackage{microtype}      
\usepackage{graphicx}
\usepackage{amsmath}
\usepackage{amssymb}
\usepackage{siunitx}
\usepackage{bbm}
\usepackage{siunitx}
\usepackage[switch]{lineno}
\usepackage[utf8]{inputenc}
\usepackage{amsmath,amsfonts,amsthm}
 \author{Neeraj Kumar\\
{Hike Private Limited}\\
neerajku@hike.in
\And
Srishti Goel\\
{Hike Private Limited}\\
srishtig@hike.in
\And
Ankur Narang\\
{Hike Private Limited}\\
ankur@hike.in
\And
Brejesh Lall \\
{IIT Delhi}\\
brejesh@ee.iitd.ac.in}

\title{Few Shot Adaptive Normalization Driven Multi-Speaker Speech Synthesis}

\begin{document}

\maketitle
\begin{abstract}
The style of the speech varies from person to person and every person exhibits his or her own style of speaking that is determined by the language, geography, culture and other factors. Style is best captured by prosody of a signal. High quality multi-speaker speech synthesis while considering prosody and in a few shot manner is an area of active research with many real-world applications. While multiple efforts have been made in this direction, it remains an interesting and challenging problem. 

	In this paper, we present a novel few shot multi-speaker speech synthesis approach (FSM-SS) that leverages adaptive normalization architecture with a non-autoregressive multi-head attention model. Given an input text and a reference speech sample of an unseen person, FSM-SS can generate speech in that person's style in a few shot manner. Additionally, we demonstrate how the affine parameters of normalization help in capturing the prosodic features such as energy and fundamental frequency in a disentangled fashion and can be used to generate morphed speech output. We demonstrate the efficacy of our proposed architecture on multi-speaker VCTK and LibriTTS datasets, using multiple quantitative metrics that measure generated speech distortion and MoS, along with speaker embedding analysis of the generated speech vs the actual speech samples.
\end{abstract}

\section{Introduction}

A lot of exciting developments have been made in speech synthesis systems to synthesize natural sounding human speech. The developments in this area have helped in a number of applications including audiobook narration, news readers, conversational assistants and engaging user experiences in the virtual worlds. 

	To realise a natural speech synthesis system, the model has to capture the speaking style of every person. For this prosodic features of speech play an important role. Prosody is a confluence of a number of phenomena such as paralinguistic information, intonation, stress, and style. Such phenomena are best described by the duration, fundamental frequency and energy of any speech. Multiple efforts are being made to incorporate and control such features into the model to capture and synthesize the speech in a person's speaking style. 

	 High quality multi-speaker speech synthesis (with prosody consideration) in a few shot manner is an interesting and challenging research problem. Present approaches for state-of-the-art TTS (Text to Speech Synthesis) such as Tacotron~\citep{tacotron}, Fast Speech~\cite{FastSpeech}, Fast speech 2~\cite{FastSpeech2} have focussed on generating the speaking style of a single speaker. These approaches do not generate audio on multiple speakers. Some of the current approaches~\citep{MultispeakTTS, MultiTTS, NeuralVoice, Deepvoice3} have used the speaker embedding to capture the identity and speaking style of the person in the speech. Such approaches fail to generate expressive speech as they have not taken the prosodic features and emotions into account and hence have lower quality in generated speech. While some of these approaches consider zero-shot approach for multi-speaker speech synthesis, none of them consider few shot explicit prosody transfer. Other approaches~\citep{expressivetacotron} rely on prosodic features such as fundamental frequency, duration and energy to generate the expressive speech. Such approaches are able to generate the expressive speech for the speakers which are the already part of training. Such approaches are not able to generate expressive speech in few shot manner on multiple speakers.

	We propose a novel approach, \textbf{FSM-SS} (Few Shot Multi-speaker Speech Synthesis), that is capable of generating speech in an unseen person’s speaking style in a few shot manner. Our model uses non-autoregressive feed forward transformer based architecture~\citep{FastSpeech2} along with adaptive normalization to generate the speech on an unseen person’s style. The model takes as inputs: $(a)$ an unseen text, and $(b)$ a reference speech sample of an unseen speaker, and generates high quality speech for the given text in the given person's speaking style. 
	
	Our main contributions are as follows:

\begin{itemize}
	\item We have proposed a novel few shot approach (FSM-SS) that uses adaptive normalization along with non-autoregressive feed forward transformer based architecture. FSM-SS can generate multi-speaker speech output in a few shot manner, given an input unseen text and an unseen person's reference speech sample.
	\item For adaptive normalization, we have proposed two architectures based on convolution and on multi-head attention to capture the prosodic properties in the network through affine parameters. This helps to capture the various affine parameters based on speaker embedding, pitch and energy.
	\item We have proposed that the affine parameters of instance normalization are able to capture the information of speaker identity, pitch and energy. Conditioning on the pitch, energy and speaker embedding generates personalized and temporally smoother speech which captures the speaking style of a person much better than known state-of-the-art approaches. 
	\item Using extensive experiments on multi-speaker VCTK and LibriTTS datasets, we show both qualitative and quantitative improvements over prior approaches along with high quality of output and the capability of our approach to generate speech for a wide variety of unseen speakers.
	\item FSM-SS can also be used as a voice morphing tool by varying the embedding, frequency and energy inputs to the adaptive normalization module.
\end{itemize}

\section{Related Work}

Earlier work in prosody and modeling of the speaking style has been studied since the era of HMM-based speech synthesis. In ~\citep{HMM}, expressive clusters are generated using hierarchical k-means clustering and then HMM-based speech synthesis is used to provide a flexible framework to model the varying expressions. In ~\citep{HMM1}, multiple emotional expressions and speaking styles of speech are modeled in a single model by using a multiple-regression hidden semi-Markov model and the authors proposed estimating the transformation matrix for a set of predefined style vectors. Our approach uses non-autoregressive deep neural networks based method instead of HMM-based speech generation.

Various efforts such as ToBI ~\citep{HMM2}, AuToBI ~\citep{HMM3}, INTSINT ~\citep{HMM4}, SLAM ~\citep{SLAM} have described methods for the annotation and automatic labeling of prosody. Such methods often require domain experts, however, and inter-rater annotations can differ substantially. Our approach uses deep learning techniques to transfer the prosodic features on generated speech instead of manual labeling of prosody.

After the advent of deep learning techniques, a lot of work has been done in text to speech generation on multiple speakers. VoiceLoop ~\citep{voiceLoop} proposed a novel architecture based on a fixed size memory buffer that can generate speech from voices unseen during training. However, obtaining good results required tens of minutes of enrollment speech and transcripts for a new speaker. ~\citep{VoiceLoop1} extended VoiceLoop to utilize a target speaker encoding network to predict a speaker embedding. This network is trained jointly with the synthesis network using a contrastive triplet loss to ensure that embeddings predicted from utterances by the same speaker are closer than embeddings computed from different speakers. In addition, a cycle-consistency loss is used to ensure that the synthesized speech encodes to a similar embedding as the adaptation utterance. Our proposed approach (FSM-SS) uses a pretrained speaker embedding model~\citep{G2E} to feed speaker embedding via adaptive normalization into a non-autoregressive architecture to generate speech for an unseen speaker.

~\citep{DeepSpeech2} introduced a multispeaker variation of Tacotron which learned low-dimensional speaker embeddings for each training speaker and phoneme durations are predicted first and then are used as inputs to the frequency model. CNN-based multispeaker model ~\citep{Deepvoice3} develops many sophisticated mechanisms in the speaker embedding and attention block to ensure the synthesized quality. These systems learn a fixed set of speaker embeddings and therefore only support synthesis of voices already seen during training. ~\citep{DeepSpeech2} and ~\citep{Deepvoice3} have used autoregressive methods to generate speaker embedding, whereas our proposed approach (FSM-SS) uses adaptive normalization along with non-autoregressive multi-head attention architecture~\citep{FastSpeech2} to generate speech, leading to faster training and inference and better quality as compared to these methods. Adaptive normalization in FSM-SS helps in few shot multi-speaker speech synthesis.

~\citep{NeuralVoice} used multi-head attention for generating speaker embedding. To see the effectiveness, they have used DeepVoice 3~\citep{Deepvoice3} TTS architecture to generate multi-speaker speech. For speaker adaptation, they have shown the few shot approach to generate speech on unseen speakers. They have used the speaker classification method which used the convolution and GRU layer to calculate the PLDA score which is then passed to the sigmoid layer. FSM-SS leverages~\citep{G2E} based $256$ speaker embedding, rather than multi head attention based $128$ speaker embedding, to generate speech on unseen speaker. The addition of pitch and energy into our proposed approach in normalisation helps in the transfer of prosodic features from the reference speech sample to generated speech.

	VAE-based method has been further leveraged~\citep{hieTTS} to handle noisy multi-speaker speech data and can control latent attributes in the generated speech that are rarely annotated in the training data, such as speaking style, accent, background noise, and recording conditions. Our proposed method uses disentangled pitch and energy of the reference speech sample to synthesize speech whearas ~\cite{hieTTS} uses probabilistic hierarchical generative model to disentangle style attributes. ~\cite{MultispeakTTS} used RNN-based Tacotron 2 that enjoys the benefits of recurrent attention computation and leverages the attention information in previous steps to help the attention calculation in the current step. They have utilized a network that is independently-trained for a speaker verification task on a large dataset of untranscribed audio from tens of thousands of speakers, using a state-of-the-art generalized end-to-end loss. ~\citep{MultiTTS} introduced a diagonal constraint on the weight matrix of the encoder-decoder attention during training and inference and employed a bottleneck structure in the decoder pre-net which encourages the decoder to generalize on the representation of speech frame instead of memorization, and forces the decoder to attend to text/phoneme inputs. Our proposed method uses adaptive normalization architecture along with  non-autoregressive multi head attention network to generate high quality speech on unseen speakers.

All the methods discussed above have either used CNN and Transformer based TTS~\citep{transformer, Deepvoice3} that can speed up the training over RNN-based models~\citep{tacotron}. All the models generate a melspectrogram conditioned on the previously generated melspectrograms and suffer from slow inference speed. These autoregressive models generate melspectrograms one by one automatically, without explicitly leveraging the alignments between text and speech. Fast Speech ~\citep{FastSpeech} speeds up the synthesis on single speaker through parallel generation of melspectogram. Fast Speech relies on the autoregressive teacher model to predict the phoneme duration and generated melspectogram for knowledge distillation. Fast Speech 2~\citep{FastSpeech2} uses the ground truth for phoneme duration prediction and incorporates other features such as pitch and energy in variance predictor for single speaker speech synthesis. Our proposed method (FSM-SS) leverages the feed-forward transformer based non-autoregressive approach along with variance adapter~\citep{FastSpeech2} but uses a novel adaptive normalization architecture to capture the reference style of an unseen speaker. This technique helps FSM-SS to deliver high quality multi-speaker output personalization in a few shot manner.

Many previous works have explicitly focussed on generating style based text to speech. ~\citep{expressivetacotron} incorporated  architectures to generate prosody embedding and speaker embedding which is combined with text encoder representation which goes to the tacotron based decoder to generate the speech. Conditioning Tacotron on this learned embedding space results in synthesized audio that matches the prosody of the reference signal with fine time detail even when the reference and synthesis speakers are different. This method uses Tacotron based autoregressive approach which is different from our proposed method which employs multi head attention based non-autoregressive method along with adaptive normalisation to capture the prosodic features. ~\citep{StyleToken} proposed “global style tokens” (GSTs), a bank of embeddings that are jointly trained within Tacotron, and learn to model a large range of acoustic expressiveness. The architecture consists of a reference encoder, style attention, style embedding, and sequence-to-sequence (Tacotron) model. We have used adaptive normalisation to capture prosidic features rather than attention netowrk to capture the style. ~\citep{StyleControlTTS} introduced the Variational Autoencoder (VAE) to an end-to-end speech synthesis model, to learn the latent representation of speaking styles in an unsupervised manner in tacotron 2 based framework. KL annealing is introduced to solve this problem instead of KL loss. We have used adaptive normalisation based architecture to capture the style features rather than variation autoencoder. ~\citep{DiverseTTS} has introduced the vector-quantized VAE (VQ-VAE), and a two-stage training approach to generate high fidelity speech samples. 
	
Our proposed method (FSM-SS) used normalisation based architecture along with multi head attention instead of VQ-VAE based tacotron architecture. ~\citep{hierarchicalTTS} aims to achieve disentangled control of each prosody attribute at different levels (utterance, word and phone levels) and proposes a multilevel model based on Tacotron 2 integrated with a hierarchical latent variable model. Our proposed method uses adaptive normalisation instead of an hierachical approach to capture the prosody.

\section{FSM-SS Design}

In this section, we present the overall design and architecture of FSM-SS including adaptive normalization and non-autoregressive multi-head attention based feed forward transformer for few shot multi-speaker speech synthesis. 

\subsection{Model Overview}

The speech synthesis model uses non-autoregressive multi-head attention feed forward transformer~\citep{FastSpeech2} which is state of the art in speech synthesis for a single speaker. It helps in parallel melspectrogram generation and speeds up the speech synthesis compared to Fast Speech and autoregressive models such as Transformer based TTS ~\citep{transformer, Deepvoice3} and Tactotron based TTS ~\citep{tacotron}. It uses multi-head attention-based encoder-decoder architecture along with the variance adapter method. 

We have designed two architectures for adaptive normalization: one based on multi-head attention network~\citep{mha} and another on a convolution network to learn the affine parameters in normalization. The inputs to the normalization module are: speaker embedding and pitch and energy values per frame extracted from the given reference speech sample, of an unseen person. The speaking style of the unseen speaker is fine-tuned on trained proposed architecture using a few audio-text pairs for few shot inference.  

This architecture is used to generate the melspectogram and the final audio is generated by using Griffin-Lim spectrogram inversion ~\citep{griffin} and Wave Glow architecture ~\citep{waveglow}.

\subsection{ Architecture}

\begin{figure}[h]
  \begin{center}
   \includegraphics[width=0.8\linewidth]{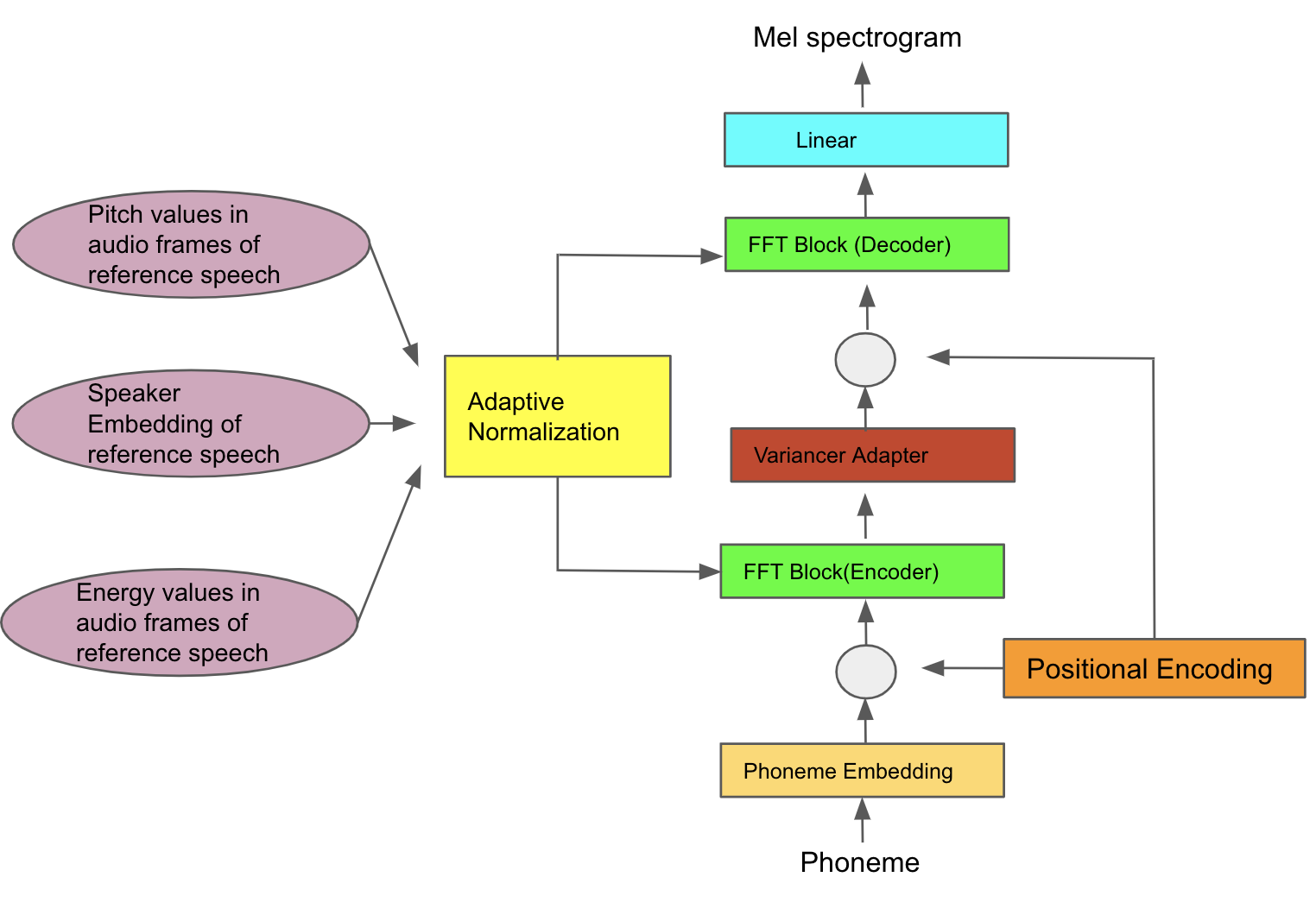}
   \caption{FSM-SS : Few Shot style based text to Speech Generation Architecture}
   \label{fig:main-arch}
  \end{center}
\end{figure}

Fig.~\ref{fig:main-arch} illustrates the architecture used in FSM-SS. During training, it takes as input: the text-audio pairs of a person along with his(her) reference speech samples. During inference, it takes a few unseen text-audio pairs along with one reference speech sample on an unseen speaker, to generate speech in that person's speaking style. The adaptive normalization is applied both during encoder and decoder stages (Fig.~\ref{fig:main-arch}) and hence helps in prosody transfer in a few shot manner. 

\paragraph{ Feed-Forward Transformer}

\begin{figure}[h]
  \begin{center}
   \includegraphics[width=0.8\linewidth]{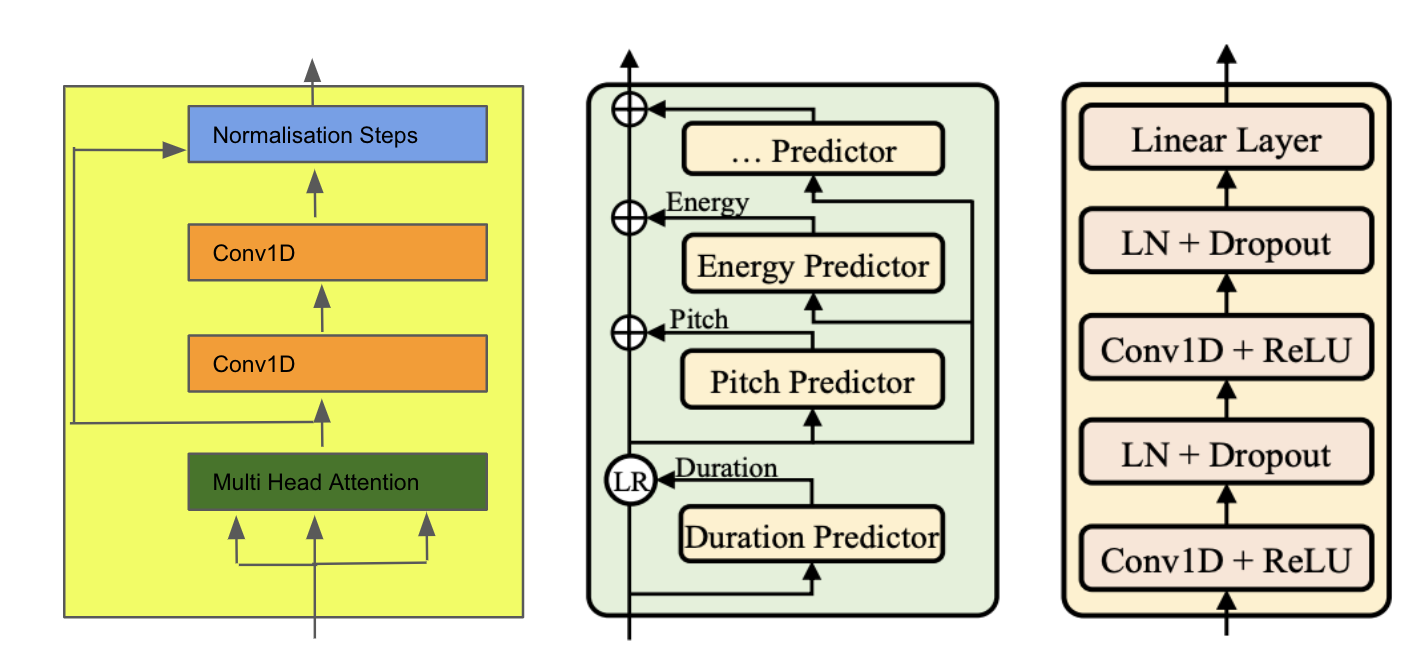}
   \caption{Left: FFT block , Centre : Variance Adaptor , Right : Variance Predictor}
   \label{fig:FS2}
  \end{center}
\end{figure}

The architecture of Feed-Forward Transformer (Fig.~\ref{fig:FS2}) is based on a multi-head self-attention network, and position feed-forward network which consists of two Conv1D and normalization stages. The proposed method stacks multiple FFT blocks with phoneme embedding and position encoding as an input as the phoneme side, and multiple FFT blocks for the melspectrogram generation, with variance adapter in between. 

\paragraph{Adaptive Normalization Stage}
This stage consists of adaptive normalization with learnable parameters such as $\gamma$ and $\beta$ which are computed through two proposed approaches: based on convolution network and multi-head attention network. This helps in adjusting the bias and scale of the normalized features to learn the required properties of speech signal including prosody. This module enables adaptive instance normalization of the feature map coming as output from the prior FFT block (Fig.~\ref{fig:FS2}).

\begin{figure}[h]
  \begin{center}
   \includegraphics[width=0.8\linewidth]{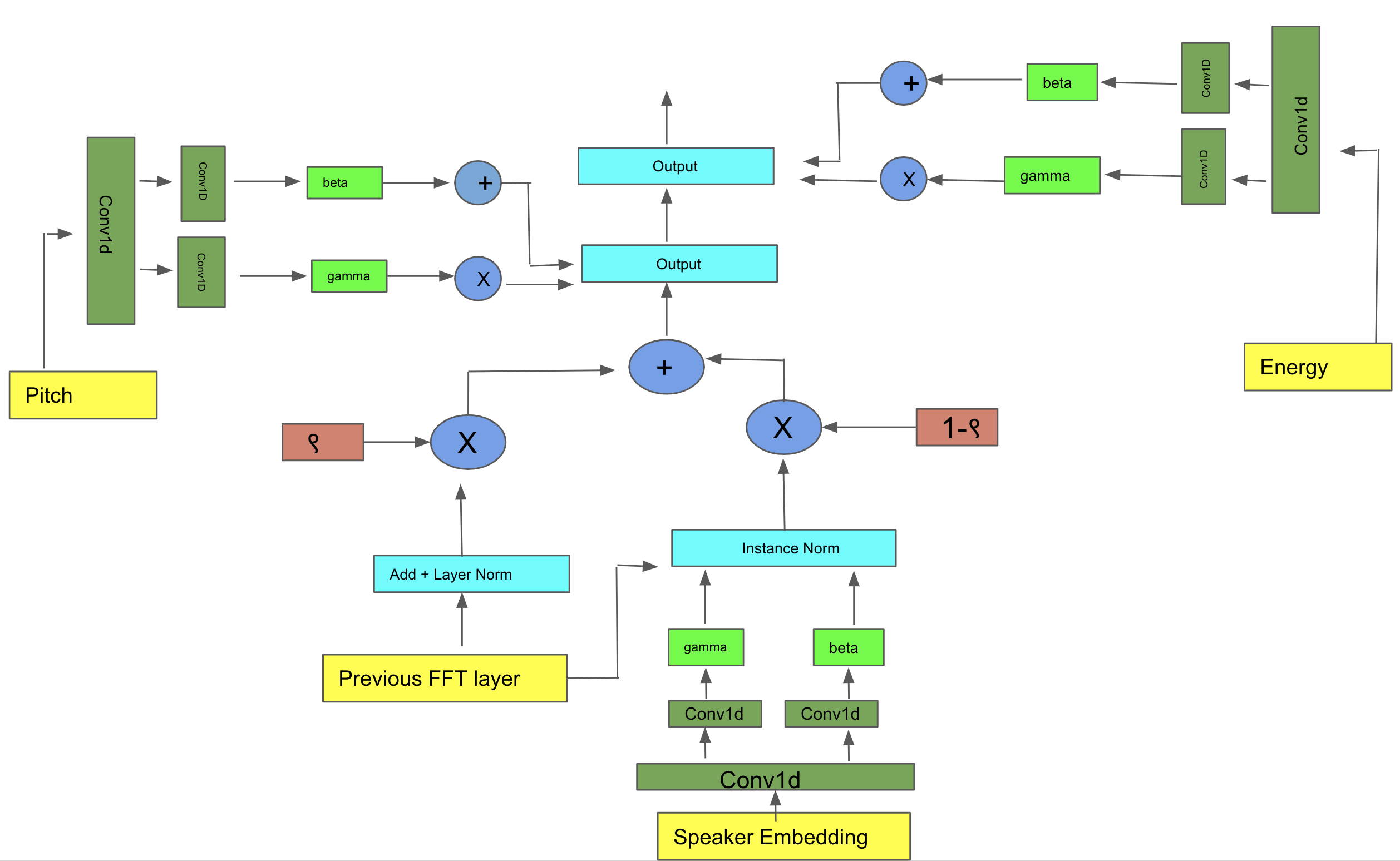}
   \caption{Convolution based normalization in proposed FSM-SS architecture}
   \label{fig:conv-norm}
  \end{center}
\end{figure}

\paragraph{Convolution based Normalization}
 We have taken three audio-related features: speaker embedding, fundamental frequency and energy of the reference speech sample which are important for capturing the prosody of reference speech. These three features are passed into the convolution layer to generate the affine parameters (Fig.~\ref{fig:conv-norm}). The parameter $\rho$ is used to combine these parameters (Equation~\eqref{conv}). The value of $\rho$'s is constrained to the range of [0, 1] simply by imposing bounds at the parameter update step. We employ a residual connection around each of the two sub-layers, x = z + Sublayer(z), followed by layer normalization,  where Sublayer(z) is the convolution function implemented by the sub-layer itself. The other part of this equation has instance normalization having  $\gamma\textsubscript{SE}$ and $\beta\textsubscript{SE}$ coming from speaker embedding. The second equation (Equation~\eqref{conv2}) generates the affine parameters from the energy and pitch values for each frame of the reference speech sample.

\begin{equation}
\label{conv}
    y = \rho(\gamma\textsubscript{LN} x\textsubscript{LN} + \beta\textsubscript{LN}) + (1- \rho)(\gamma\textsubscript{SE} x\textsubscript{IN} + \beta\textsubscript{SE})
\end{equation}

\begin{equation}
\label{conv2}
\gamma\textsubscript{energy}. (\gamma\textsubscript{pitch}. y + \beta\textsubscript{pitch} ) + \beta\textsubscript{energy} 
\end{equation}
\vspace{-1.0em}

\paragraph{Multi Head Attention Based Normalization}

\begin{figure}[h]
  \begin{center}
   \includegraphics[width=0.8\linewidth]{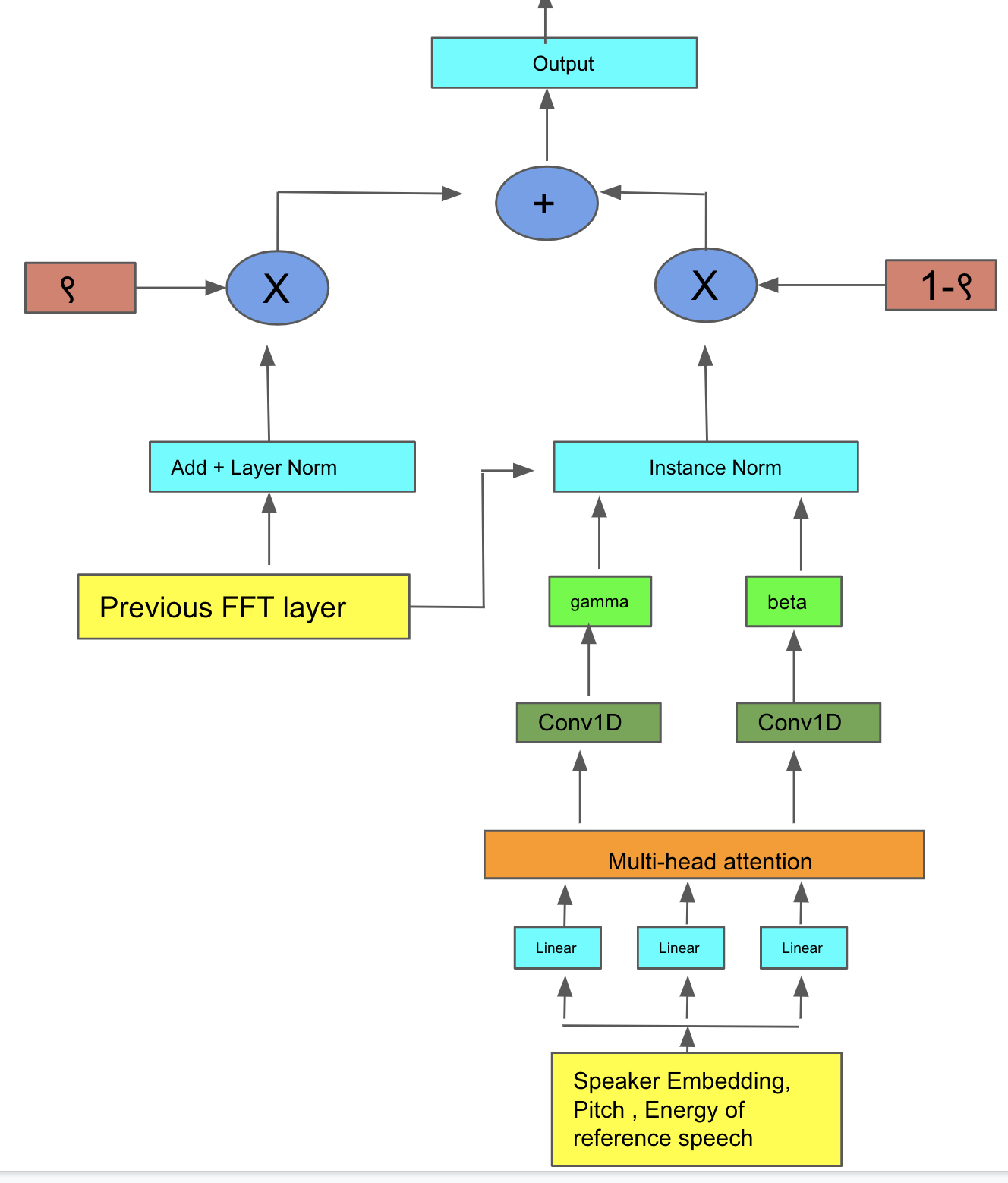}
   \caption{Multi Head Attention Based Normalization in proposed FSM-SS architecture}
   \label{fig:norm-attn}
  \end{center}
\end{figure}

In this architecture (Fig.~\ref{fig:norm-attn}), we have concatenated the speaker embedding ($256$ dimensional vector), frequency and energy of the reference speech sample to generate a tensor of size (audio-frames * 258 * batches). This is then fed it into the multi-head attention network to generate the affine parameters. These affine parameters (Equation~\eqref{attn}) are used to bias $\beta\textsubscript{attention}$ and scale $\gamma\textsubscript{attention}$ the output feature map coming from previous FFT block (Fig.~\ref{fig:FS2}).

\begin{equation}
\label{attn}
     \rho(\gamma\textsubscript{LN} x\textsubscript{LN} + \beta\textsubscript{LN}) + (1- \rho)(\gamma\textsubscript{attention} x\textsubscript{IN} + \beta\textsubscript{attention})
\end{equation}

The speaker embeddings, frequency and energy are concatenated and passed to the linear layer independently to become query, key and values of multi-head attention layer. The multi-head attention equation is given by: 

\begin{equation}
     Attention(Q,K,V) = softmax(\dfrac{QK\textsuperscript{T}}{d\textsubscript{k}})V
\end{equation}

\paragraph{Variance Adapter}

The variance predictor is used to predict the prosodic features of speech such as duration, fundamental frequency and energy. The variance adapter consists of three predictors namely: the duration predictor, pitch predictor and energy predictor. During the training phase, all three predictors are trained with the ground truth of duration, pitch and energy through three separate variance predictors independently and optimized with mean square error. 

\paragraph{Variance Predictor}

Variance predictor consists of a 2-layer 1D-convolution network with ReLU activation, each followed by the layer normalization and the dropout layer, and an extra linear layer to project the hidden states into the output sequence~\citep{FastSpeech2}. For the duration predictor, the output is the length of each phoneme in the logarithmic scale. For pitch and energy predictor, the output is the frame-level fundamental frequency and energy of melspectrogram respectively.

\subsection{Few shot approach for style adaptation}
We have used few shot approach for speaker adaptation (at inference time) using the reference speech sample of an unseen person and text. At inference, we update the whole model on a few samples of unseen speech and text pairs, while the reference speech sample remains the same since that provides prosody information via adaptive normalization. Training the whole model with all the losses gives more degrees of freedom. Early stopping is used to avoid overfitting.

\section{Experiments}

\subsection{Implementation Details}

\paragraph{Datasets } We train and evaluate the model on two datasets namely VCTK ~\citep{vctk} and LibriTTS multi-speaker dataset \citep{librispeech}. We have used 44 hours of speech with 108 speakers of the VCTK dataset and  586 hours of speech with 2456 speakers of the LibriTTS dataset.
\vspace{-1.0em}
\paragraph{Preprocesing Steps} To alleviate the mispronunciation problem, we convert the text sequence into the phoneme sequence ~\citep{DeepSpeech2,  tacotron} using open-source grapheme-to phoneme tool ~\citep{g2p}.  We extract the phoneme duration with MFA ~\citep{mfa}, an open-source system for speech-text alignment to improve the alignment accuracy.  

We transfer the raw waveform into melspectrograms by setting the frame size and hop size to 1024 and 256 with respect to the sample rate of 22050 Hz. We extract fundamental frequency, F0 from the raw waveform with the same hop size to obtain the pitch of each frame and compute the L2-norm of the amplitude of each STFT frame as the energy. We feed the values of pitch and energy values in the proposed normalization method.

In the training process, we quantize the F0 and energy of each frame to 256 possible values and encode them into a sequence of one-hot vectors as p and e respectively. We feed the pitch and energy embedding with p and e at the variance adapter stage. The output of pitch and energy predictors are values of F0 and energy which is minimized with mean square error.

We have generated the speaker embedding from pretrained model, Generalized end-to-end loss for speaker verification ~\citep{G2E}  which is trained on : (1) LibriSpeech Other ~\citep{librispeech}, which contains 461 hours of speech from a set of 1,166 speakers disjoint from those in the clean subsets, (2) VoxCeleb ~\citep{voxceleb}, and (3) VoxCeleb2 ~\citep{voxceleb2} which 139K utterances from 1,211 speakers, and 1.09M utterances from 5,994 speakers, respectively.
\vspace{-1.0em}
\paragraph{Model Details} We have used $4$ feed forward transformer blocks at the phoneme encoding stage and at the output mel-spectrogram decoder stage. The dimension of phoneme embedding and hidden layer of self attention is set to 256 in every FFT block. The number of attention heads is set to $2$. The output linear layer converts the 256-dimensional hidden states into 80-dimensional mel-spectrograms. The size of the phoneme vocabulary is $76$, including punctuations.

The Convolution based normalization architecture feeds $256$ dimensional speaker embedding into 1D convolution layer to generate affine parameters. The fundamental frequency and energy of the reference speech are fed to 1D convolution layers each to reduce the channel length from max frames of speech signal in the dataset to $512$. The affine parameters are then calculated by adding 1D convolution layer to generate $256$ channel output respectively.

In multi head attention based normalization architecture, the $256$ dimensional speaker embedding is replicated along the time frame of the mel spectrogram and then concatenated with frequency and energy features to generate $258$ dimensional feature vectors for all time steps (audio frames). It is then fed to multi head attention with the number of heads set to $6$. The generated feature map is then fed to 1D convolution to generate $256$ channel output which is added with output of layer normalization using the learnable parameter $\rho$.

The Variance predictor consists of $2$ blocks of Conv1D, relu, layer normalization and dropout layer. The kernel sizes of the 1D-convolution is set to $3$, with input/output sizes of $256$/$256$ for both layers and the dropout rate is set to $0.5$.

The pretrained Wave Glow architecture ~\citep{waveglow}is used as a vocoder to generate the speech at 22050 Hz. It is trained on LibriSpeech dataset at the sampling frequency of 22050 Hz.

\paragraph{Training and Inference} 
We have used the batch size of $96$ and $64$ for the convolution-based normalization method and multi-head attention based normalization technique in the proposed architecture respectively with the initial value of $\rho$ is $0.7$. The Adam optimizer is used with $\beta 1$ = $0.9$ , $\beta 2$ = $0.98$ , $\epsilon$ =10e-9. It takes around 120K steps for the convolution-based normalization method on VCTK and LibriTTS dataset. The multi-head attention normalization based model takes 470K steps and 800K steps to converge on VCTK ~\citep{vctk} and LibriTTS datasets \citep{librispeech}. We have trained the model on 4 V100 GPU based machine. Note that the length of the reference speech sample and the speech generated from unrelated text input can be different

During inference time, we have used a few shot approach with samples from $0$ to $5$ to generate the speech in the speaking style of the reference person. We have used Wave Glow vocoder to generate the final speech from melspectrogram.

\subsection{Implementation Results}

\begin{figure}[h!]
  \begin{center}
   \includegraphics[width=0.5\linewidth]{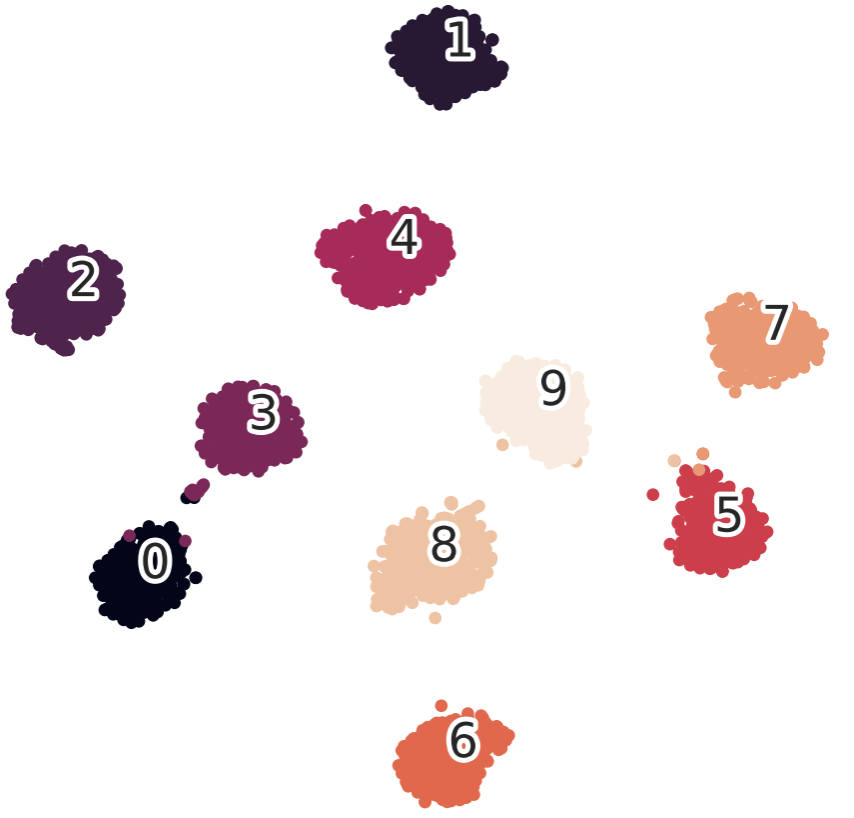}
   \caption{t-SNE visualization of speaker embeddings of generated samples of VCTK dataset. Cluster id 0 to 4 refers to male speakers and 5 to 9 refers to female speakers}
   \label{fig:pca_male_female}
  \end{center}
\end{figure}

\begin{figure}[h!]
  \begin{center}
   \includegraphics[width=0.8\linewidth]{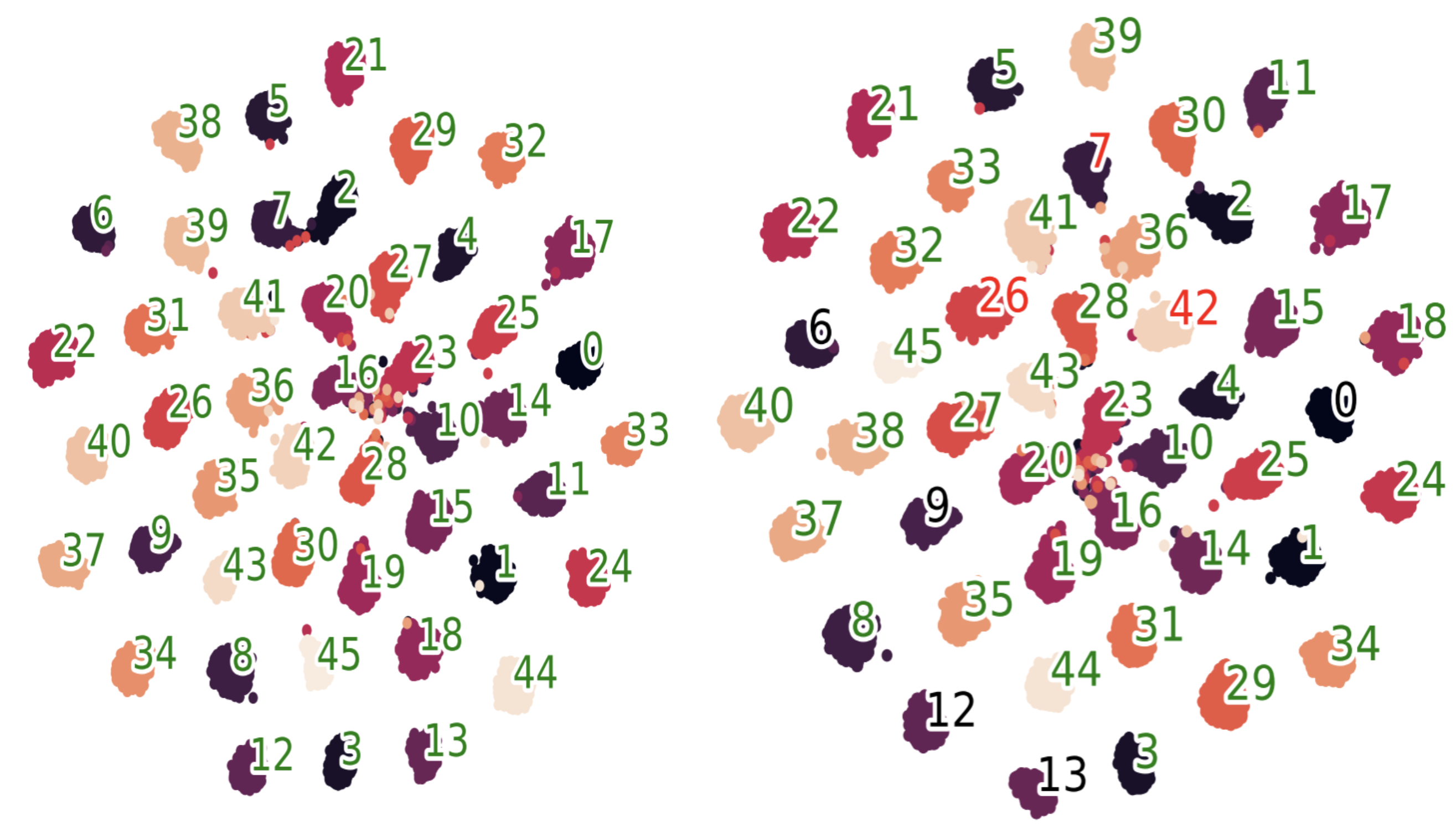}
   \caption{t-SNE Visualization of speaker embeddings of male actual and generated samples of VCTK dataset. The left side shows the actual embedding space of all male speakers in the dataset. Right side shows the embedding space of train(green), val(red: cluster id - 7,26,42) and test(black: cluster id - 0,6,9,12,13) }
   \label{fig:embedd}
  \end{center}
\end{figure}

\paragraph{ Speaker Embedding Space}
The speech samples are generated on test speakers to visualize how well different samples are spread on embedded space. We have generated the 256 dimensional embedding of every speech and done the t-SNE visualization which shows that the synthesized utterances on the same speaker tend to lie very close in the embedding space, demonstrating the consistency of generation. The visualization is done on the speech synthesized in zero-shot approach on an unseen speaker. Figure ~\ref{fig:pca_male_female} shows that generated embedding on male and female speakers form distinct clusters. In Figure ~\ref{fig:embedd} we have done the t-SNE visualization where we have shown that we are able to generate the samples from speakers that are far away from the clusters correctly, which demonstrates the variety of multiple unseen speakers that can be handled by our approach.

\begin{figure}[h]
  \begin{center}
   \includegraphics[width=0.8\linewidth]{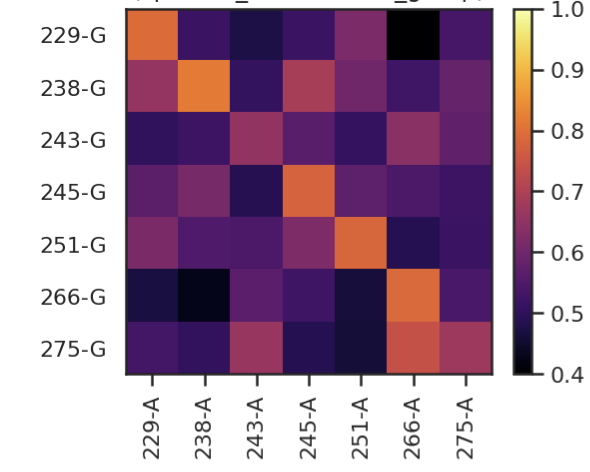}
   \caption{Cross Similarity between utterances of actual speaker(x-axis) and generated speaker(y-axis) of VCTK dataset}
   \label{fig:cross_sim}
  \end{center}
\end{figure}

\begin{figure}[h]
  \begin{center}
   \includegraphics[width=0.5\linewidth]{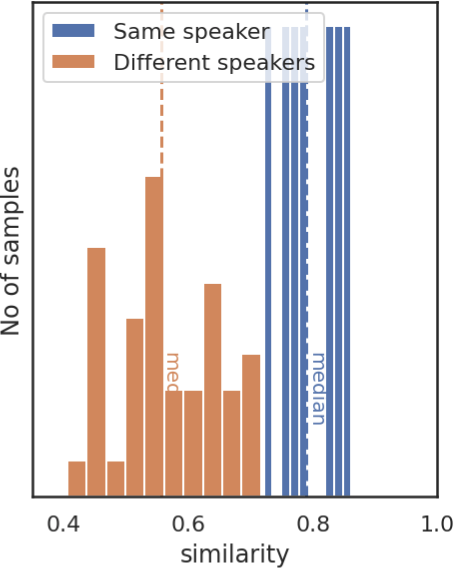}
   \caption{Normalized histogram of similarity values between utterances actual speaker and generated speaker pf VCTK dataset}
   \label{fig:histogram}
  \end{center}
\end{figure}

\paragraph{Speaker Similarity}
We expect the utterances from the same speaker to have high similarity values and those distinct to have lower one. We have evaluated the cosine similarity as the similarity metric on the speaker embedding of generated samples with actual samples. We have extracted the speech in zero shot approach on unseen speaker. Figure ~\ref{fig:cross_sim} shows that the higher similarity of emebedding on actual and generated speech for same speaker. Figure ~\ref{fig:histogram} shows that the median values of cosine similarities are higher for same speaker and lower for different speaker.

\begin{figure}[h!]
  \begin{center}
   \includegraphics[width=0.6\linewidth]{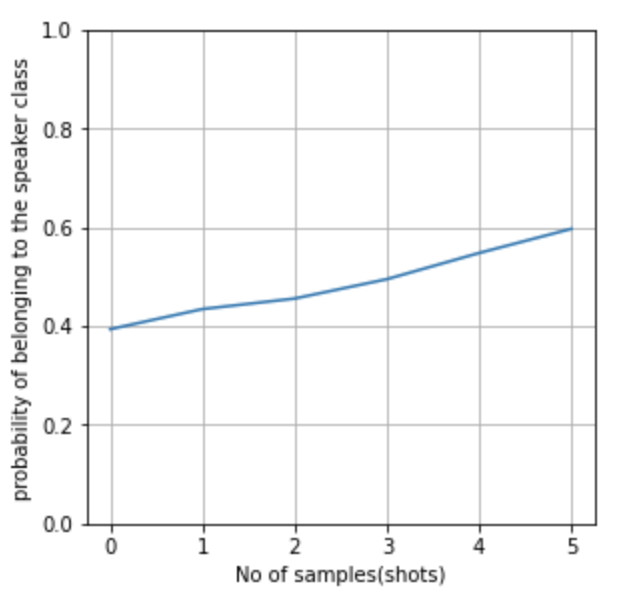}
   \caption{probability of belonging to a speaker class on few show approach with different number of unseen speaker samples and text pairs.  }
   \label{fig:classfication}
  \end{center}
\end{figure}

\vspace{-1.0em}
\paragraph{Speaker Classification on few shot approach}
We have used the few shot approach for speaker adaptation by providing different audio and text pair of unseen speaker. We have used gaussian naive bayes multilabel classifier ~\citep{nbc} whose accuracy is around $92\%$. Figure ~\ref{fig:classfication} shows that with increase in the number of samples in few shot approach the  probability of speaker identification has increased from from $0.39$ to $0.59$. ~\citep{NeuralVoice} has shown improvement to the probability of $0.55$ when 5 samples are used for few shot approach.  

\paragraph{Audio Quality}

Twenty samples of speakers with different accents are taken for VCTK test and Twenty samples of english speaking speakers from LibriTTS are used to perform mean opinion score. The text content is kept consistent among different systems so that all testers only examine the audio quality without other interference factors. Table~\ref{tab:quantita} shows better MOS score than NVS ~\citep{NeuralVoice} as they have used 128 dimensional speaker embedding based on multi head attention network with transformer based TTS architecture to generate sample, whereas, FSM-SS uses~\citep{G2E} based $256$ speaker embedding pretrained on 3 datasets along with pitch and energy values for speech synthesis. 

\begin{table}[h!]
  \begin{center}

    \begin{tabular}{c|c|c} 
      \textbf{Method} & \textbf{VCTK$\uparrow$} & \textbf{LibriTTS$\uparrow$}\\
      \hline
       GT   & \textbf{4.05 $\pm$ 0.05}& \textbf{ 4.10$\pm$ 0.24 } \\
       GTmel+waveglow  & 3.84 $\pm$ 0.14 & 3.92 $\pm$ 0.46 \\
       Conv+waveglow  & 3.75 $\pm$ 0.56  & 3.45 $\pm$ 0.68  \\
       Attention+waveglow & 3.72 $\pm$ 0.24 & 3.38 $\pm$ 0.08  \\
       NVS& 3.13 $\pm$ 0.42 &  -   \\
    \end{tabular}
    \vspace {0.25\baselineskip}
    \caption{MOS score on FSM-SS with 95$\%$ confidence interval for VCTK and LibriTTS dataset. GT - ground Truth , GTmel+waveglow - groung truth mel spectrogram with waveglow as vocoder, Conv+waveglow - Convolution based normalsation in FSM-SS with waveglow vocoder , Attention+waveglow - Multi head attention based normalisation in FSM-SS with waveglow architecture, NVS : Neural Voice cloning with few samples method }
    \label{tab:quantita}
    
  \end{center}
\end{table}

Apart from subjective evaluation, we have used the metrics namely Gross Pitch Error ~\citep{vde}, Voicing Decision Error ~\citep{vde}, F0 Frame Error ~\citep{ffe}, Mel Cepstral Distortion ~\citep{mcd} which are used in audio signal processing to measure the prosody of the signal. The qualitative and quantitative metrics are extracted on the speech generated in the zero-shot approach on unseen speakers. The generated outputs from FSM-SS architecture are given in \footnote{ Generated audios : \url{https://sites.google.com/view/fsmss/home}}

 Table~\ref{tab:quantitative} shows that the convolution-based normalization has lower errors compared to multi-head attention based normalization. Table~\ref{tab:quantitative1} shows that the few-shot approach is able to lower the errors than the zero-shot approach. ~\citep{expressivetacotron} has higher  MCD(10.87) due to the use of tacotron based encoders to capture the pitch and speaker embedding whereas FSM-SS uses pitch, energy and speaker embedding of reference speech though proposed normalization methods. ~\citep{hierarchicalTTS} have shown lower MCD(8.8) on LibriTTS dataset on seen speakers due to multi-resolution architecture of prosody while FSM-SS has an MCD value of 9.78 on unseen speakers.

\begin{table}[h!]
  \begin{center}

    \begin{tabular}{c|c|c|c|c}
      \textbf{Method} & \textbf{MCD$\downarrow$} & \textbf{GPE$\downarrow$} & \textbf{VDE$\downarrow$} &  \textbf{FFE$\downarrow$} \\
      \hline
       Conv-1   & \textbf{13.65}& \textbf{28.70} &  \textbf{18.04} & \textbf{35.46}\\
       Attention-1  & 14.63 & 30.45 &  19.60  & 37.64\\
       Conv-2  & \textbf{14.15} & \textbf{30.50} & \textbf{19.98}  & \textbf{38.26}  \\
       Attention-2 & 15.17 & 32.51 & 21.06 & 41.64 \\
    \end{tabular}
    \vspace {0.25\baselineskip}
    \caption{Quantitative metrics on FSM-SS on zero shot approach. Conv: Convolution based normalization in FSM-SS , Attention : Multi head attention based normalization in FSM-SS, 1- VCTK dataset, 2- LibriTTS dataset}
    \label{tab:quantitative}
    
  \end{center}
\end{table}
\vspace{-1.0em}
\begin{table}[h!]
  \begin{center}

    \begin{tabular}{c|c|c|c|c}
      \textbf{Method} & \textbf{MCD$\downarrow$} & \textbf{GPE$\downarrow$} & \textbf{VDE$\downarrow$} &  \textbf{FFE$\downarrow$} \\
      \hline
       Conv-1   & \textbf{09.78}& \textbf{24.45} &  \textbf{14.47} & \textbf{28.90}\\
       Attention-1  & 10.56 & 26.45 &  16.12  & 30.36\\
       Conv-2  & \textbf{11.05} & \textbf{26.76} & \textbf{16.67}  & \textbf{30.58}  \\
       Attention-2 & 12.71 & 27.52 & 17.23 & 31.62 \\
    \end{tabular}
    \vspace {0.25\baselineskip}
    \caption{Quantitative metrics on FSM-SS on few shot approach (5 samples). Conv: Convolution based normalization, Attention : Multi head attention based normalization, 1- VCTK dataset, 2- LibriTTS dataset}
    \label{tab:quantitative1}
    
  \end{center}
\end{table}
\vspace{-1.0em}
\subsection{Ablation Study}
We have done the ablation study on Convolution based normalization architecture in FSM-SS with a zero-shot approach on the VCTK dataset. The base model with pitch and energy in the normalization stage without speaker embedding of reference speech has not shown very good results as the information of speaker identity is missing in the architecture. We have then used speaker embedding in the normalization steps with the base model and do not incorporate pitch and energy values of the reference unseen speaker. The quality of output degrades as the variance predictor is not able to predict the required duration, frequency, and energy values. The addition of pitch values along with speaker embedding of reference speech helps in improving the speech quality. Table~\ref{tab:table5} shows the decreasing values of different errors when adding pitch and energy in the normalization architecture.

\begin{table}[h!]
  \begin{center}
    \begin{tabular}{c|c|c|c|c|c} 
      \textbf{Method} & \textbf{MCD$\downarrow$} & \textbf{GPE$\downarrow$} & \textbf{VDE$\downarrow$} &  \textbf{FFE$\downarrow$} &\textbf{MoS$\uparrow$}\\
      \hline
       BM+P+E   & 21.68& 40.45 &  45.76 & 65.87&2.64$\pm$0.14\\
       BM+SE   & 20.65& 38.71 &  39.56 & 56.64&2.92$\pm$0.26\\
       BM+SE+P  & 16.38 & 33.65 &  29.76  & 45.76&3.32$\pm$0.08\\
       FSM-SS  & \textbf{13.65}& \textbf{28.70} &  \textbf{18.04} & \textbf{35.46}& \textbf{3.72$\pm$0.24}\\
       
    \end{tabular}
    \vspace {0.25\baselineskip}
    \caption{Ablation Study of FSM-SS. BM is Base Model without normalization method. SE is speaker embedding in normalization, P and E are pitch and energy values in the normalization network. }
    \label{tab:table5}
  \end{center}
\end{table}
\vspace{-0.25em}
\subsection{Extension of Proposed Method}

\paragraph{Voice morphing}

\begin{figure}[h]
  \begin{center}
   \includegraphics[width=0.8\linewidth]{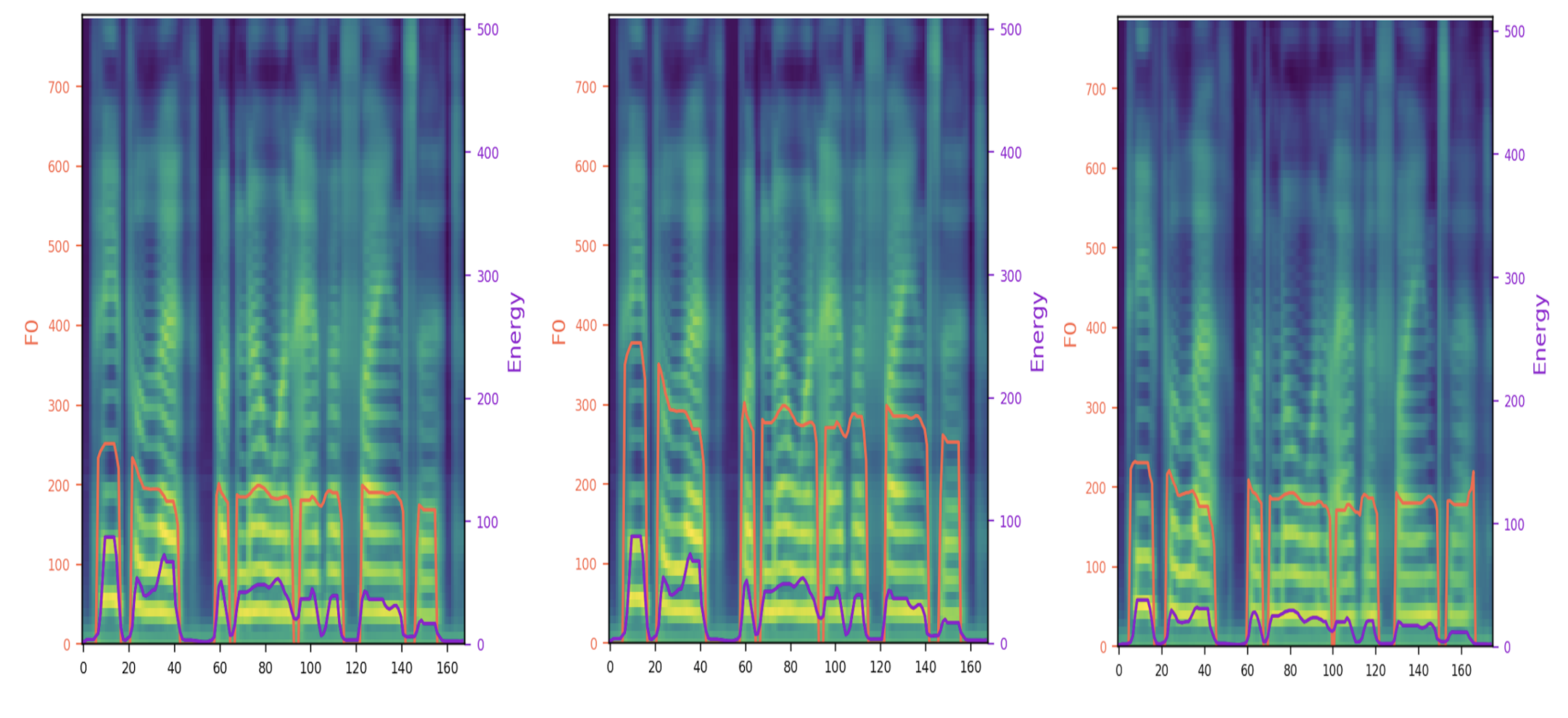}
   \caption{Left: Synthesized samples on a reference speaker and text, Centre: Pitch is modulated by increasing the F0 to 1.25F0 keeping the energy values constant on same reference speaker and text Right: Energy values are reduced from E to 0.5E keeping the pitch values constant on same reference speaker and text}
   \label{fig:morphing}
  \end{center}
\end{figure}

We can independently tune the speaker embedding, fundamental frequency and energy of the reference speech which are fed into the normalization steps to generate the morphed speech. Figure ~\ref{fig:morphing} shows that independently modulating the pitch and energy values leads to the voice morphing. This has a lot of applications in the virtual world, the gaming industry, voice modulation, etc.

\vspace{-1.0em}
\section{Conclusions}
In this paper, we have proposed a novel few shot approach (FSM-SS) that uses adaptive normalization along with non-autoregressive feed forward transformer based architecture. FSM-SS can generate multi-speaker speech output in a few shot manner, given an input unseen text and an unseen person's reference speech sample. For adaptive normalization, we have proposed two architectures based on convolution and on multi-head attention to capture the prosodic properties in the network through affine parameters. This helps to capture the various affine parameters based on speaker embedding, pitch and energy. Using extensive experiments on multi-speaker VCTK and LibriTTS datasets, we show both qualitative and quantitative improvements over prior approaches along with high quality of output and the capability of our approach to generate speech for a wide variety of unseen speakers. FSM-SS can also be used as a voice morphing tool by varying the embedding, frequency and energy inputs to the adaptive normalization module.



\end{document}


\maketitle

\section{Evaluation Metric}
We have used the metrics namely Gross Pitch Error ~\citep{vde}, Voicing Decision Error ~\citep{vde}, F0 Frame Error ~\citep{ffe}, Mel Cepstral Distortion ~\citep{mcd} which are used in audio signal processing to measure the prosody of the signal.All pitch and voicing metrics are computed using the output of the YIN ~\citep{Yin} pitch tracking algorithm.
For all comparisons of predicted signals to target signals, we extend the shorter signal to the length of the longer signal using a domain-appropriate padding  (0 for a time domain waveform and for a log magnitude spectrogram with a $1*10\textsuperscript{-6}$  stabilizing offset)

\begin{itemize}
\item Gross Pitch Error ~\citep{vde}  - the percentage of frames for which the absolute pitch error is higher than a certain threshold. For speech, this threshold is usually 20\% where p\textsubscript{t},$p^{\prime}\textsubscript{t}$ are the pitch signals from the reference and predicted audio, v\textsubscript{t},$v^{\prime}\textsubscript{t}$ are the voicing decisions from the reference and predicted audio, and $\mathbbm{1}$ is the indicator function.

\begin{align*}
        GPE = \dfrac{\sum_{t}\mathbbm{1}[|p\textsubscript{t} - p^{\prime}\textsubscript{t}| - 0.2p\textsubscript{t}]\mathbbm{1}[v\textsubscript{t}]\mathbbm{1}[v^{\prime}\textsubscript{t}]}{\sum_{t}\mathbbm{1}[v\textsubscript{t}][v^{\prime}\textsubscript{t}]}
\end{align*}

\item Voicing Decision Error ~\citep{vde}  - The percentage of frames for which an incorrect voiced/unvoiced decision is made  where v\textsubscript{t},$v^{\prime}\textsubscript{t}$ are the voicing decisions from the reference and predicted audio, and $\mathbbm{1}$ is the indicator function and T is the total number of frames.

\begin{align*}
        VDE = \dfrac{\sum_{t=0}^{t=T-1}\mathbbm{1}[v\textsubscript{t}\neq v^{\prime}\textsubscript{t}]}{T}
\end{align*}
\item F0 Frame Error ~\citep{ffe} - percentage of frames where either a GPE or VDE is observed.
\begin{align*}
        FFE = \dfrac{\sum_{t}\mathbbm{1}[|p\textsubscript{t} - p^{\prime}\textsubscript{t}| - 0.2p\textsubscript{t}]\mathbbm{1}[v\textsubscript{t}]\mathbbm{1}[v^{\prime}\textsubscript{t}] + \mathbbm{1}[v\textsubscript{t}\neq v^{\prime}\textsubscript{t}]}{T}
\end{align*}

 \item Mel Cepstral Distortion ~\citep{mcd} - It is the measure of how different two sequences of mel cepstra are. It is used in assessing the quality of parametric speech synthesis systems Where c\textsubscript{t}, $c^{\prime}\textsubscript{t}$ are the k-th mel frequency cepstral coefficient (MFCC) of the t-th frame from the reference and predicted audio. We sum the squared differences over the first K MFCCs

\begin{equation}
    MCD\textsubscript{K} = \dfrac{1}{T} \sum_{t=0}^{T-1}\sqrt{\sum_{k=1}^{K}(c^{\prime}\textsubscript{t,k} - c\textsubscript{t,k})}
\end{equation}

\end{itemize}

The generated outputs from FSM-SS architecture are given in \footnote{ Generated audios : \url{https://sites.google.com/view/fsmss/home}}

\section{Model Hyperparameters}

\begin{table}[h!]
  \begin{center}

    \begin{tabular}{c|c} 
      \textbf{Hyperparameter} & \textbf{FSM-SS}\\
      \hline
       Phoneme Embedding Dimension & 256 \\
       encoder layer & 4 \\
       encoder attention head & 2 \\
       encoder hidden & 256 \\
       decoder layer & 4 \\
       decoder attention head & 2 \\
       decoder hidden & 2 \\
       Encoder/Decoder conv1d filter size & 1024 \\
       Encoder/Decoder Conv1D Kernel size & 9 \\
       Encoder/Decoder Dropout & 0.2 \\
       variance predictor filter size & 256 \\
       variance predictor kernel size & 3 \\
       variance predictor dropout & 0.5 \\
       
    \end{tabular}
    \vspace {0.25\baselineskip}
    \caption{Hyperparameters of FSM-SS }
    \label{tab:quantita}
    
  \end{center}
\end{table}

\begin{table}[h!]
  \begin{center}

    \begin{tabular}{c|c} 
      \textbf{Hyperparameter} & \textbf{Conv}\\
      \hline
       Pitch embedding  & 1 \\
       frequency embedding & 1 \\
       speaker embedding(SE) & 256 \\
       conv1d kernel size & 9 \\
       Pitch/frequency intial filter conv1d filter size & 512 \\
       Pitch/frequency gamma and beta conv1d filter size & 256 \\
       Speaker embedding intial filter conv1d filter size & 512 \\
       Speaker embedding gamma and beta conv1d filter size & 256 \\

    \end{tabular}
    \vspace {0.25\baselineskip}
    \caption{Hyperparameters of Convolution based normalization  architecture of FSM-SS }
    \label{tab:quantita}
    
  \end{center}
\end{table}

\begin{table}[h!]
  \begin{center}

    \begin{tabular}{c|c} 
      \textbf{Hyperparameter} & \textbf{Attention}\\
      \hline
       Pitch embedding  & 1 \\
       frequency embedding & 1 \\
       speaker embedding(SE) & 256 \\
       Concatenated pitch/frequency/SE embedding & 258 \\
       conv1d kernel size & 9 \\
       attention head & 6 \\
       gamma and beta conv1d filter size & 256 \\

    \end{tabular}
    \vspace {0.25\baselineskip}
    \caption{Hyperparameters Multi head attention based Normalization architecture of FSM-SS }
    \label{tab:quantita}
    
  \end{center}
\end{table}

\section{Experimental Results}

We have computed the evaluation metrics for few shot approach on VCTK dataset. Table~\ref{tab:eve}  shows that the metrics have decreased with the increasing number of shots on Convolution based normalization in FSM-SS method..  
\begin{table}[h!]
  \begin{center}

    \begin{tabular}{c|c|c|c|c} 
      \textbf{Shots} & \textbf{MCD$\downarrow$} & \textbf{GPE$\downarrow$} & \textbf{VDE$\downarrow$} &  \textbf{FFE$\downarrow$}\\
      \hline
       1-shot   & 13.42& 27.56 &  17.52 & 34.44\\
       2-shot   & 11.78& 27.08 &  17.18 & 33.38 \\
       3-shot  &  10.91 &  26.04 &  16.13 & 32.41 \\
       4-shot  & 10.12 & 25.34 & 15.71 & 30.56 \\
       5-shot  & 09.78 & 24.45 & 14.47 & 28.90 \\
    \end{tabular}
    \vspace {0.25\baselineskip}
    \caption{Metric for few shot approach for convolution based normalization in FSM-SS for VCTK dataset  }
    \label{tab:eve}
  \end{center}
\end{table}

Table~\ref{tab:attentioeve} shows that the few shot approach for attention based normalization in FSM-SS has improved the evaluation metrics. The errors have decreased with the increase in the number of shots.

\begin{table}[h!]
  \begin{center}

    \begin{tabular}{c|c|c|c|c} 
      \textbf{Shots} & \textbf{MCD$\downarrow$} & \textbf{GPE$\downarrow$} & \textbf{VDE$\downarrow$} &  \textbf{FFE$\downarrow$}\\
      \hline
       1-shot   &14.12 & 29.44 &  19.39 & 37.11\\
       2-shot   & 13.39& 28.66 &  18.06 & 35.38 \\
       3-shot  &  12.09 & 26.89 &  18.22 & 33.74 \\
       4-shot  & 11.77 & 26.03 & 17.21 & 32.91 \\
       5-shot  & 11.18 & 25.91 & 16.12 & 31.23 \\
    \end{tabular}
    \vspace {0.25\baselineskip}
    \caption{Metric for few shot approach for attention based normalization in FSM-SS for VCTK dataset  }
    \label{tab:attentioeve}
  \end{center}
\end{table}

We have computed the evaluation metrics to see the effect of ratio of reference speech length with input speech length in zero shot approach on convolution based normalization in FSM-SS method. Table~\ref{tab:eve1} shows that the The range of ratio (0.8 to 1.3) have lowest metrics as compared to other ratio range. The errors have increased for higher and lower ratios range. All the range have lower metrics and the speech generated are of good quality.

\begin{table}[h!]
  \begin{center}

    \begin{tabular}{c|c|c|c|c}  
      \textbf{Ratio-range} & \textbf{MCD$\downarrow$} & \textbf{GPE$\downarrow$} & \textbf{VDE$\downarrow$} &  \textbf{FFE$\downarrow$}\\
      \hline
       0.3-0.8   & 14.54& 29.45 &  19.82 & 37.17\\
       0.8-1.3   & 11.53& 26.73 &  16.14 & 33.91 \\
       1.3-1.7  &  13.97 &  29.31 &  18.18 & 36.71 \\
       1.7-2.3  & 15.31 & 30.34 & 20.59 & 39.47 \\
       2.3-5  & 16.51 & 30.96 & 17.16 & 38.79 \\
       5-8 &   16.71 & 30.51 & 17.14 & 38.09 \\
    \end{tabular}
    \vspace {0.25\baselineskip}
    \caption{Metric for zero shot approach for convolution based normalization in FSM-SS for VCTK dataset  .Ratio-range denotes the range of ratio of  reference speech length with input speech length  }
    \label{tab:eve1}
  \end{center}
\end{table}

\section{Ablation Study}
We have done the ablation study for few show approach for VCTK dataset on convolution architecture of FSM-SS . The errors are decreasing with the increase in the number of shots.

\begin{table}[h!]
  \begin{center}

    \begin{tabular}{c|c|c|c|c}  
      \textbf{Method} & \textbf{MCD$\downarrow$} & \textbf{GPE$\downarrow$} & \textbf{VDE$\downarrow$} &  \textbf{FFE$\downarrow$}\\
      \hline
       BM+P+E   & 20.91& 40.04 &  44.56 & 65.46 \\
       BM+SE   & 19.78& 37.98 &  39.90 & 56.05 \\
       BM+SE+P  & 16.20 & 33.68 &  29.94  & 45.72 \\
       FSM-SS  & \textbf{13.42}& \textbf{27.56} &  \textbf{17.52} & \textbf{34.44}\\
    \end{tabular}
    \vspace {0.25\baselineskip}
    \caption{Ablation Study of FSM-SS on 1 shot approach }
    \label{tab:table5}
  \end{center}
\end{table}

\begin{table}[h!]
  \begin{center}

    \begin{tabular}{c|c|c|c|c} 
      \textbf{Method} & \textbf{MCD$\downarrow$} & \textbf{GPE$\downarrow$} & \textbf{VDE$\downarrow$} &  \textbf{FFE$\downarrow$} \\
      \hline
       BM+P+E   & 19.51 & 39.84 &  43.52 & 62.34 \\
       BM+SE   & 19.20& 36.34 &  39.42 & 54.42 \\
       BM+SE+P  & 14.38 & 31.22 &  20.86  & 40.24 \\
       FSM-SS  & \textbf{11.78}& \textbf{27.08} &  \textbf{17.18} & \textbf{33.38} \\
    \end{tabular}
    \vspace {0.25\baselineskip}
    \caption{Ablation Study of FSM-SS on 2 shot approach }
    \label{tab:table5}
  \end{center}
\end{table}

\begin{table}[h!]
  \begin{center}

    \begin{tabular}{c|c|c|c|c}  
      \textbf{Method} & \textbf{MCD$\downarrow$} & \textbf{GPE$\downarrow$} & \textbf{VDE$\downarrow$} &  \textbf{FFE$\downarrow$} \\
      \hline
       BM+P+E   & 18.78& 39.12 &  42.88 & 61.31 \\
       BM+SE   & 18.04& 35.45 &  37.89 & 51.45 \\
       BM+SE+P  & 13.77 & 29.43 &  18.91  & 37.65 \\
       FSM-SS  & \textbf{10.91}& \textbf{26.04} &  \textbf{16.13} & \textbf{32.92} \\
       
    \end{tabular}
    \vspace {0.25\baselineskip}
    \caption{Ablation Study of FSM-SS on 3 shot approach }
    \label{tab:table5}
  \end{center}
\end{table}

\begin{table}[h!]
  \begin{center}

    \begin{tabular}{c|c|c|c|c} 
      \textbf{Method} & \textbf{MCD$\downarrow$} & \textbf{GPE$\downarrow$} & \textbf{VDE$\downarrow$} &  \textbf{FFE$\downarrow$} \\
      \hline
       BM+P+E   & 17.11& 37.37 &  40.85 & 56.43 \\
       BM+SE   & 16.49& 33.86 &  36.43 & 46.32 \\
       BM+SE+P  & 13.01 & 28.65 &  18.76  & 35.76\\
       FSM-SS  & \textbf{10.12}& \textbf{25.34} &  \textbf{15.71} & \textbf{30.56} \\
       
    \end{tabular}
    \vspace {0.25\baselineskip}
    \caption{Ablation Study of FSM-SS on 4 shot approach }
    \label{tab:table5}
  \end{center}
\end{table}

a
\begin{table}[h!]
  \begin{center}

    \begin{tabular}{c|c|c|c|c} 
      \textbf{Method} & \textbf{MCD$\downarrow$} & \textbf{GPE$\downarrow$} & \textbf{VDE$\downarrow$} &  \textbf{FFE$\downarrow$} \\
      \hline
       BM+P+E   & 16.43& 37.67 &  39.42 & 53.87 \\ 
       BM+SE   & 15.85& 32.69 &  34.61 & 44.59 \\
       BM+SE+P  & 12.38 & 26.79 &  16.02  & 31.45 \\
       FSM-SS  & \textbf{09.78}& \textbf{24.45} &  \textbf{14.47} & \textbf{28.90} \\
       
    \end{tabular}
    \vspace {0.25\baselineskip}
    \caption{Ablation Study of FSM-SS on 5 shot approach }
    \label{tab:table5}
  \end{center}
\end{table}

\section{Evaluation text speaker}

\subsection{Test speaker for VCTK dataset}

Table~\ref{tab:vctk} shows the test speakers of VCTK dataset ~\citep{vctk}.

\begin{table}[h!]
  \begin{center}

    \begin{tabular}{c|c|c|c|c}  
      \textbf{Speaker ID } & \textbf{Gender} & \textbf{Nationality} \\
      \hline
       229   & Female & english \\
       238  & Female & NorthernIrish \\
       266   & Female & Irish \\
       228  & Female  & English \\
       231  & Femle  & English \\
       288  & Female & Irish  \\
       243  & Male  & English \\
       245  & Male & irish \\
       251  & Male & Indian \\
       275  & Male & Scottish \\
       273  & Male  & English \\
       
    \end{tabular}
    \vspace {0.25\baselineskip}
    \caption{Test Speakers from VCTK dataset }
    \label{tab:vctk}
  \end{center}
\end{table}

\subsection{Test speaker for LibriTTS dataset}
Table~\ref{tab:libritts} shows the test speakers of LibriTTS dataset \citep{librispeech}.

\begin{table}[h!]
  \begin{center}

    \begin{tabular}{c|c|c|c|c} 
      \textbf{Speaker ID } & \textbf{Gender} \\
      \hline
       6829   & Female \\
       9026  & Female \\
       8975   & Female \\
       6696  & Female  \\
       192  & Femle  \\
       557  & Male \\
       1355  & Male  \\
       176  & Male \\
       3144  & Male \\
       4345  & Male \\
       1065  & Male \\
       
    \end{tabular}
    \vspace {0.25\baselineskip}
    \caption{Test Speakers from LibriTTS dataset }
    \label{tab:libritts}
  \end{center}
\end{table}

\section{MOS Interface}

\begin{figure}[h!]
  \begin{center}
   \includegraphics[width=0.8\linewidth]{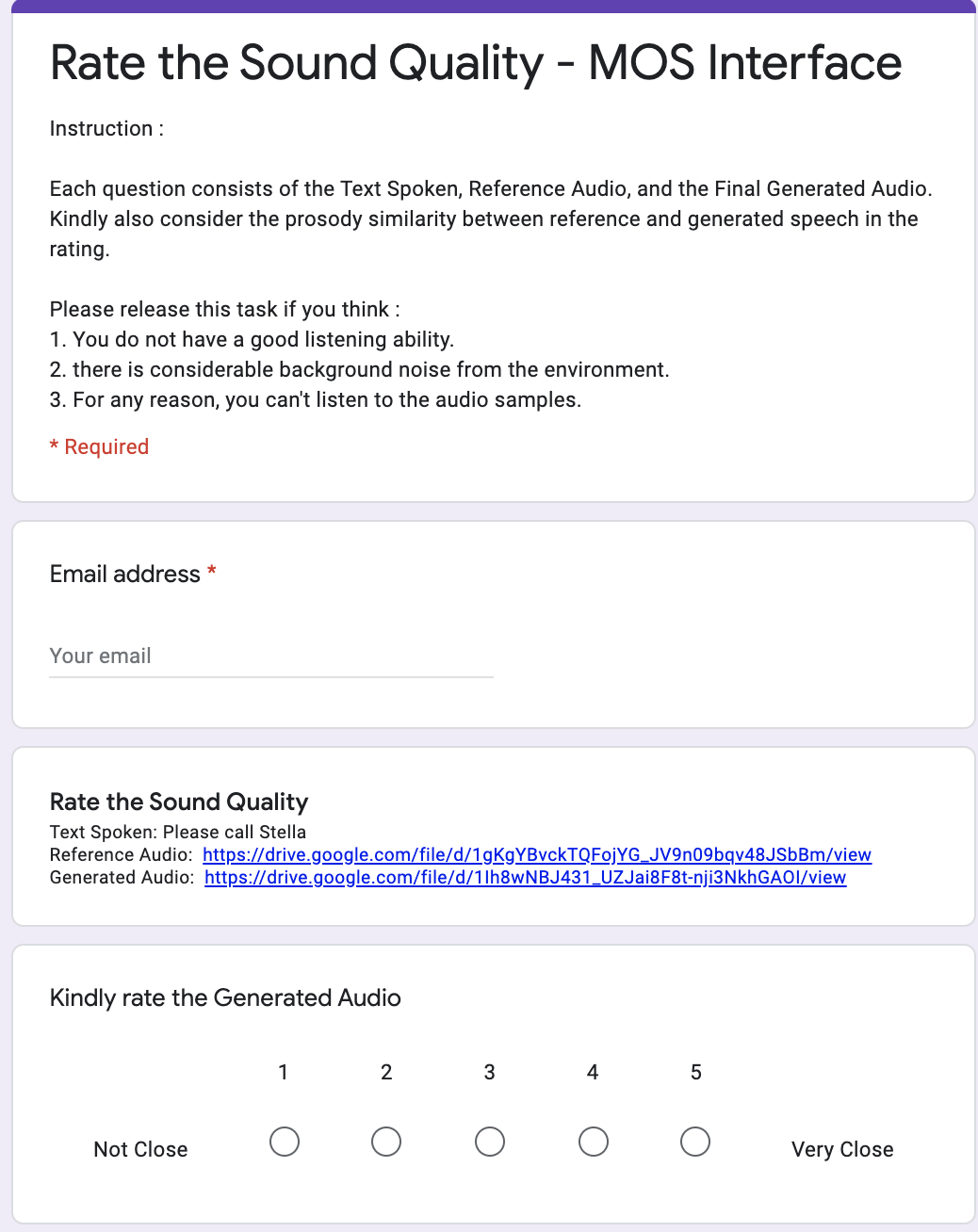}
   \caption{MOS interface for calculating the MOS score of the proposed method.}
   \label{fig:MOS}
  \end{center}
\end{figure}

Figure ~\ref{fig:MOS} shows the MOS interface to rate the quality of speech generated from proposed FSM-SS.

